\documentclass[aps,prl,twocolumn,groupedaddress]{revtex4}

\usepackage{graphicx}

\begin{document}

\renewcommand{\vec}[1]{{\mathbf #1}}

\title{Photocount statistics in mesoscopic optics}
\author{S. Balog}
\affiliation{Department of Physics, University of Fribourg, 1700
Fribourg, Switzerland}
\author{P. Zakharov}
\affiliation{Department of Physics, University of Fribourg, 1700
Fribourg, Switzerland}
\author{F. Scheffold}
\email[]{Frank.Scheffold@unifr.ch}
\affiliation{Department of
Physics, University of Fribourg, 1700 Fribourg, Switzerland}
\author{S.E. Skipetrov}
\email[]{Sergey.Skipetrov@grenoble.cnrs.fr}
\affiliation{Laboratoire de Physique et Mod\'elisation des Milieux Condens\'es/CNRS,\\
Maison des Magist\`{e}res, Universit\'{e} Joseph Fourier, 38042
Grenoble, France}

\date{\today}

\begin{abstract}
We report the first observation of the impact of mesoscopic fluctuations on the photocount statistics
of coherent light scattered in a random medium. Poisson photocount distribution of the incident light
widens and gains additional asymmetry upon transmission through a suspension of small dielectric spheres.
The effect is only appreciable when the average number $\overline{n}$ of photocounts
becomes comparable or larger than the effective dimensionless conductance $g$ of the sample.
\end{abstract}


\maketitle


Since the Anderson's discovery that the propagation of a quantum
particle can be blocked by disorder \cite{anderson58} and
subsequent realization that this `Anderson localization' can also
take place for electromagnetic waves or photons (light)
\cite{john84}, the quest for observing it has become a
breathtaking adventure
\cite{wiersma97,scheffold99,gomez99,chabanov00}. Although the
observation of microwave localization in quasi-one dimensional
disordered samples \cite{chabanov00} now seems to be accepted by
the scientific community, the localization of visible light in
strongly scattering, three-dimensional (3D) semiconductor powders
\cite{wiersma97} has been questioned \cite{scheffold99}. The
extensive subsequent work \cite{gomez99} has shown that the
scattering strength of available disordered materials in the
optical frequency range is not sufficient to clearly distinguish
the impact of Anderson localization on the commonly measured
quantities (such as the average transmission coefficient and the
coherent backscattering cone) from the impact that would be
produced by a weak absorption of light in diffuse regime. For this
reason, well-established experimental results in the field of
optical localization in 3D media are limited to `precursors' of
Anderson localization that can be observed under conditions of
diffuse scattering: coherent backscattering and weak localization
\cite{albada85}, long-range correlations
\cite{albada90,scheffold97}, and universal conductance
fluctuations \cite{scheffold97}.

The mesoscopic optical phenomena that we cited above can be
understood and discussed in the framework of classical physics as
of 1905, without appealing to quantum mechanics. The impact of
quantum-mechanical effects on the coherent backscattering of light
has been demonstrated in beautiful experiments on light scattering
in cold atomic clouds \cite{labeyrie99}. The quantum nature of
\textit{scatterers\/} (atoms) had to be taken into account to
understand the low value of the coherent backscattering
enhancement factor. Experimental studies revealing the quantum
nature of \textit{light\/} in multiple scattering have been
reported only very recently \cite{lodahl05}, besides a considerable
theoretical interest in this subject \cite{patra99}.
In particular, Kindermann \textit{et al.} have predicted that disorder can
substantially alter photon statistics of degenerate \textit{incoherent\/}
radiation.
In the present Letter we report the first experimental observation of the
impact of one of the precursors of Anderson localization ---
mesoscopic, long-range correlations
--- on the photon statistics of degenerate \textit{coherent\/} light
emitted by a conventional continuous laser.
We interpret our results in the framework of the semiclassical theory
of photoelectric detection \cite{mandel95}, which appears to be sufficient under
conditions of experiments reported here.

As first noted by Einstein 100 years ago \cite{einstein05}, the
quantum nature of light can be directly probed by the
photo-electric effect. As the energy of the electromagnetic wave
is quantized in portions $\hbar \omega$ (with $\hbar$ the Planck
constant and $\omega$ the frequency of light),
only an integer number $n$ of such quanta (photons) can be
absorbed by a photo-electric effect-based detector during a given
time interval $\tau$. Today's electronic equipment allows us to
measure $n$ in a wide dynamic range and to determine the
probability distribution of photocounts $P(n, {\overline{n}})$, where
${\overline{n}}$ is the average number of photocounts in the interval
$\tau = {\overline{n}}/f$ and $f$ is the average photocount rate (number
of counts per unit time).
$P(n, \overline{n})$ carries fundamental information about interaction of light
with the medium.
In a `random laser', for example, $P(n, \overline{n})$ can be used to characterize
different regimes of lasing \cite{cao01}.
In a different domain of physics ---
mesoscopic electronics --- the statistics of quasi-particle
(electron) counts (`full counting statistics') in disordered conductors
is also under active study \cite{levitov96}.

\begin{figure}
\includegraphics[height=5cm,angle=0]{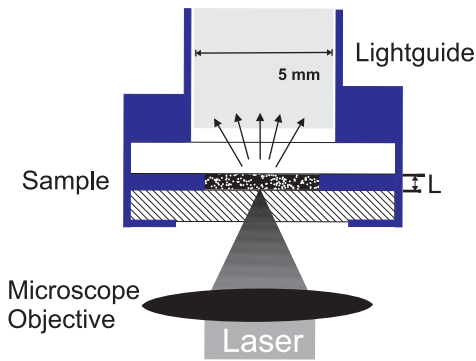}
\caption{\label{fig1} Experimental setup: The incident laser beam
is focused on the vertically oriented sample holder. The glass
window directed towards the laser is highly absorbing while the
opposite window is transparent. The windows hold a sealed liquid
layer (thickness $L$) of colloidal titanium dioxide suspended in
water. Light transmitted through the sample is collected by a
light guide and recorded by a single photon detector (not shown).}
\end{figure}

\begin{figure}
\includegraphics[width=8.3cm,angle=0]{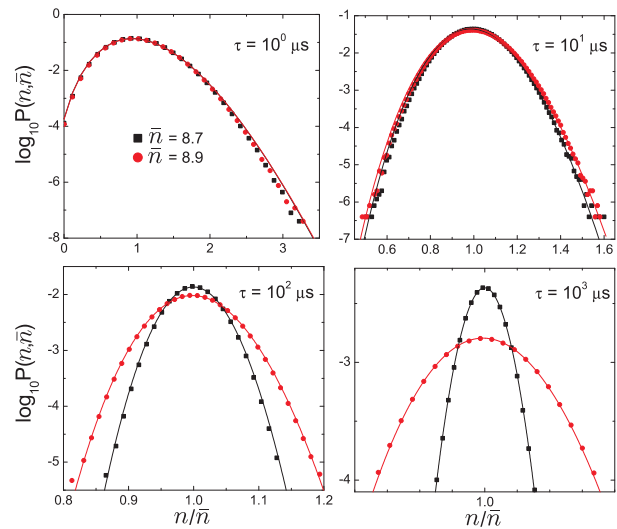}
\vspace*{-0.3cm}\caption{\label{fig2} Probability distributions of
photon counts for a wide incident beam (black squares) and a
focused beam (beam waist $w = 3.4$ $\mu$m, red circles) for four
different sampling times $\tau$. The former distribution follows
the Poisson law (black line), while the latter one is well
described by the Fourier transform of our Eq.\ (\ref{mandel}) (red
line). Small deviations for the shortest $\tau$ can be explained
by the finite detector dead time \cite{johnson66}. The number of
data points for the two longest $\tau$ has been reduced to improve
readability.}
\end{figure}

\begin{figure}
\includegraphics[width=7.7cm,angle=0]{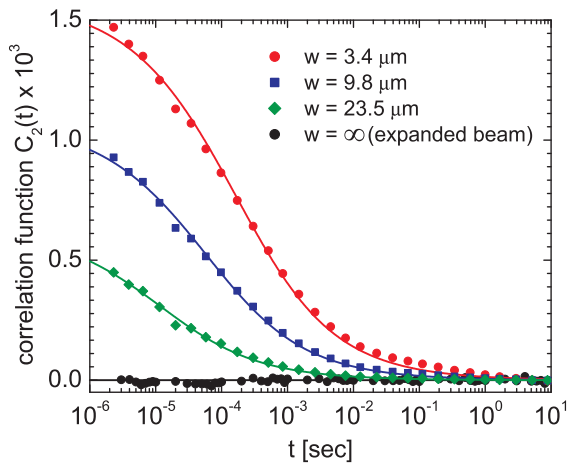}
\caption{\label{fig3} Correlation function of total transmission
$T$ for different beam spot sizes $w$ (distance between $1/e$
intensity values of a focused Gaussian beam). Lines are fits to
the data.}
\end{figure}

In our experiment we measure the distribution of photocounts of
laser light transmitted through an optically dense slab (see Fig.\
\ref{fig1}).  The sample cell (thickness $L \approx 0.5$ mm) is
filled with a charge stabilized aqueous colloidal dispersion of a
commercial titanium dioxide powder (Warner Jenkinson Europe Ltd.),
particle diameter $\approx$ 200--300 nm, at an initial density of
$18\pm 1 \%$ per volume. To further increase the density we let
the sealed suspension settle under gravity. Due to the
electrostatic repulsion between the particles the sedimentation is
asymptotically slowed down and an equilibrium layer of approx.
0.2--0.25 mm thickness (volume fraction ca. 35--40\%) is formed
after about 10 hours. From diffusing-wave spectroscopy
measurements we have checked that the particles in this layer
remain mobile and undergo Brownian motion. Following the approach
described in reference \cite{scheffold97} we estimate the
transport mean free path at this density to $l^* = 0.7 \pm 0.1$
$\mu$m. Our sample is therefore deep in the multiple scattering
regime: a typical transmitted photon experiences $(L/l^*)^2 \sim
10^5$ elastic scattering events, whereas the coherent incident
beam is destroyed after a distance $l^* \ll L$ and hence does not
contribute to the measured signal. The glass window directed
towards the laser is highly absorbing (transmission coefficient
$\simeq 0.001$) in order to suppress multiple reflections
\cite{scheffold97} while the opposite one is transparent. A
frequency doubled Nd:YV04 laser (`Verdi' from Coherent) operating
at the wavelength $\lambda = 532$ nm illuminates the sample
through a microscope objective that focuses the laser beam to a
small spot (spot size $w \gtrsim 3$ $\mu$m) on the sample surface.
The light transmitted through the cell is collected by a light
guide (core diameter 5 mm), recorded by a single photon detector
with short dead time, and processed by a digital photon counter
(correlator.com, New Jersey, USA). The high temporal resolution
(12.5 ns) of the latter assures that no more than one photon is
arriving every time step for a typical photon count rate $f$ of 9
MHz. In a typical experiment the photon trace is recorded over one
hour. From the recorded data we compute the probability
distribution $P(n, {\overline{n}})$ and the time averaged
correlation function of total transmission $C_2(t)=\langle T(0)
T(t) \rangle / {\overline{T}}^2-1$ following standard procedures
\cite{schatzel90}. To suppress contributions from slow drifts the
data is analyzed in thirty second intervals and subsequently
averaged. Some representative results for different beam spot
sizes are shown in Figs.\ \ref{fig2} and \ref{fig3}.

Because the detection process is of probabilistic nature, the
detection of a photon is a random event and $P(n, {\overline{n}})$
is expected to be the Poisson probability distribution
\cite{mandel95}. However, as follows from Fig.\ \ref{fig2}, this
is only true when the incident laser beam is sufficiently wide.
For a focused beam we observe that the distribution widens and
becomes more asymmetric than one would expect for the Poisson
distribution. This indicates that additional fluctuations exist in
the latter case. These additional fluctuations are due to the
random motion of scatterers. Since we collect all the transmitted
intensity (total transmission measurements), and since the surface
of our sample is much larger than the typical size $\sim \lambda$
of speckle spots, one could naively expect spatial speckle to
average out, as it is indeed the case for large $w$. However, if
$w$ is small and the scattering is sufficiently strong, coherent
interferences of scattered light give rise to weak but long-range
correlation of distant speckle spots, which acquire a synchronous
component in their fluctuations. This results in enhanced
fluctuations of the total transmitted signal \cite{scheffold97}.

The enhanced fluctuations of the total transmission $T$ can be
studied assuming that light is a classical wave described by
Maxwell equations \cite{deBoer94,rossum99}. As long as
localization effects are weak, the statistical distribution
$P_T(T)$ of $T$ appears to be very close to Gaussian (because the
many speckle spots contributing to $T$ are only weakly correlated
and the central limit theorem applies), but with enhanced second
and non-zero third central moments $M_T^{(2)}$ and $M_T^{(3)}$,
where $M_T^{(k)} = \langle (T - {\overline{T}})^k
\rangle/{\overline{T}}^k$. Therefore, the characteristic function
of $T$ can be approximated by
\begin{eqnarray}
\chi_T(q) = \exp\left( i {\overline{T}} q - \frac{1}{2} {\overline{T}}^2 M_T^{(2)} q^2 -
\frac{i}{6} {\overline{T}}^3 M_T^{(3)} q^3 \right)
\label{char}
\end{eqnarray}
When the measurement of $T$ is not instantaneous but involves time integration,
the second moment is given by \cite{mandel95}
\begin{eqnarray}
M_T^{(2)} = \frac{2}{\tau} \int\limits_0^\tau \left( 1-
\frac{t}{\tau} \right) C_2(t) dt \label{m2}
\end{eqnarray}
The third moment $M_T^{(3)}$ of the distribution of $T$ can be
shown to be proportional to the square of the second one:
$M_T^{(3)} = \alpha M_T^{(2)2}$, where the proportionality
constant $\alpha = 16/5$ for a wide ($w \gg L$) Gaussian beam
\cite{rossum99}. The limit of wide beam has also been studied in
the previous correlation experiments \cite{scheffold97} and the
corresponding correlation function $C_2(t)$ has been analyzed
theoretically \cite{berk94}. In our experiments, on the contrary,
the beam width $w$ is much smaller than the thickness $L$ of the
sample (typically, $w/L \sim 10^{-2}$). In this situation, by
performing calculations similar to that of Ref.\ \cite{berk94} we
find $C_2(t) = (2/3 g) \exp(3 t/4 t_0)[1 - \Phi(\sqrt{3 t/4
t_0})]$, where $\Phi$ is the error function. Leaving the
discussion of the microscopic expressions for $g$ and $t_0$ for a
future publication \cite{sandor06}, we just note here
that $g$ scales roughly as $1/w$ and hence the magnitude of the total transmission
fluctuations ($\sim 1/g$) can be varied by adjusting the beam spot size $w$.
By analogy with the case of disordered waveguide \cite{chabanov00},
we will further term $g$ the `effective' dimensionless conductance.
The above expression for $C_2(t)$
with $g \sim 10^3$ and $t_0 \sim 10^{-5}$--$10^{-4}$ s provides a good fit to
our measurements (see Fig.\ \ref{fig3}).

\begin{figure}
\includegraphics[width=7.7cm,angle=0]{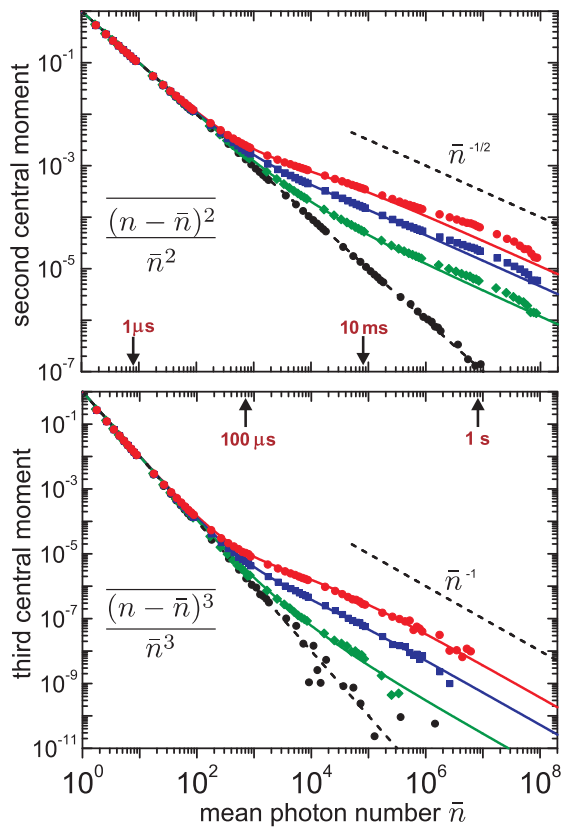}
\caption{\label{fig4} Second and third central moments of the
photocount distribution $P(n, \overline{n})$ for the same beam
spot sizes as in Fig.\ \ref{fig3}. Lines are theoretical results
(\ref{moment2}) and (\ref{moment3}). The third moment plot is a
fit to the data with $\alpha =$ 3.02, 2.59 and 2.00 (for
curves from top to bottom).}
\end{figure}

According to the famous Mandel's formula \cite{mandel95},
the statistical distribution of photocounts $P(n, {\overline{n}})$
can be obtained by averaging the Poisson distribution
$P_{\mathrm{Poisson}}(n, {\overline{n}} = \eta T)$ over the distribution $P_T(T)$ of the total transmission $T$,
with $\eta$ the quantum efficiency of the photodetector.
The Fourier transform of the Mandel's formula with respect to $n$ yields a relation
between the characteristic functions $\chi_n(q)$ and $\chi_T(q)$:
\begin{eqnarray}
\chi_n(q) &=& \chi_T\left[ i \overline{n} (1 - e^{i q})/\overline{T}  \right]
\nonumber \\
&\simeq& \exp \left( i {\overline{n}} q - \frac{1}{2} {\overline{n}}^2 M_n^{(2)} q^2 -
\frac{i}{6} {\overline{n}}^3 M_n^{(3)} q^3 \right)
\label{mandel}
\end{eqnarray}
where the second line is obtained by expanding the first one in power series
in $q$, which is justified for large $\overline{n}$. The second and the third central
moments of $n$ in Eq.\ (\ref{mandel}) are
\begin{eqnarray}
M_n^{(2)} &=& \frac{1}{{\overline{n}}} + M_T^{(2)}
\label{moment2}
\\
M_n^{(3)} &=&
\frac{1}{{\overline{n}}^2} + M_T^{(3)} + 3 \frac{M_T^{(2)}}{{\overline{n}}}
\label{moment3}
\end{eqnarray}
The Fano factor $F = (\langle n^2 \rangle -
\overline{n}^2)/\overline{n} = 1 + {\overline{n}} M_T^{(2)} > 1$,
which indicates photon bunching. The photocount distribution $P(n,
\overline{n})$ obtained by the Fourier transform of Eq.\
(\ref{mandel}) describes our measurements very well (see Fig.\
\ref{fig4}), when we use the fits to the correlation function
$C_2(t)$ obtained independently (Fig.\ \ref{fig3}) to determine
$M_T^{(2)}$ and $M_T^{(3)}$.

The first terms on the right-hand sides of Eqs.\ (\ref{moment2})
and (\ref{moment3}) correspond to the results expected for Poisson
distribution of photocounts. As we show in Fig.\ \ref{fig4}, these
results are recovered for small ${\overline{n}} \ll g$, when the
photocount distribution is dominated by the shot noise due to the
discreteness of $n$ and is not sensitive to the randomness of the
scattering medium. Deviations from the Poisson-like behavior start
to become important when ${\overline{n}}$ becomes comparable to
the effective dimensionless conductance $g$. The long-range
character of the correlation function $C_2(t) \sim 1/\sqrt{t}$ is
responsible for new scaling laws $M_n^{(2)} \sim
1/\sqrt{\overline{n}}$ and $M_n^{(3)} \sim 1/{\overline{n}}$ in
the limit of large ${\overline{n}}$. This behavior is well
confirmed by our measurements: as can be seen in Fig.\ \ref{fig4},
the data points indeed follow the $1/\sqrt{\overline{n}}$ and
$1/{\overline{n}}$ asymptotes shown by dashed lines \footnote{We
attribute the slight curvature of experimental curves in Fig.\
\ref{fig4} to slow fluctuations and drifts of the laser intensity
on a time scale of the order of minutes. Though already extremely
small in our experiments, these effects cannot be completely
suppressed.}. Mesoscopic fluctuations of the total transmission
$T$ due to the random motion of scatterers in the disordered
sample become dominant in this regime, whereas the shot noise is
negligible. We see therefore that the transition between small-
and large-${\overline{n}}$ regimes in $P(n, {\overline{n}})$ is
governed by localization effects, the strength of the latter being
measured by the dimensionless conductance $g$.

In conclusion, mesoscopic fluctuations of coherent light transmission
through a random medium produce measurable deviations of photocount distribution
$P(n, \overline{n})$ from Poisson law, provided that
the average number of photocounts $\overline{n}$ is comparable or larger than the effective dimensionless conductance $g$ of the random sample. This provides a new tool for studying
mesoscopic phenomena in random media, a tool that should be particularly valuable in the search
for Anderson localization of light. An interesting continuation of this work would be to
analyze $P(n, \overline{n})$
for incident light in thermal or in \textit{non-classical} (Fock, squeezed, etc.) states.

This work was financially supported by the Swiss National Science
Foundation (project No. 200021-101620). S.E.S. acknowledges
financial support of INTAS through a Young Scientist Fellowship at
the early stage of this work.



\begin{thebibliography}{99}

\bibitem{anderson58}
P.W. Anderson, \textit{Phys. Rev.} \textbf{109}, 1492 (1958).

\bibitem{john84}
S. John,
\textit{Phys. Rev. Lett.} \textbf{53,} 2169 (1984);
P.W. Anderson,
\textit{Phil. Mag. B} \textbf{52,} 505 (1985);
S. John,
\textit{Phys. Today} \textbf{44}(5), 32 (1991).

\bibitem{wiersma97}
D.S. Wiersma, P. Bartolini, A. Lagendijk, and R. Righini,
\textit{Nature} \textbf{390,} 671 (1997).

\bibitem{scheffold99}
Comment on Ref.\ \cite{wiersma97}: F. Scheffold, R. Lenke, R.
Tweer, and G. Maret, \textit{Nature} \textbf{398,} 206 (1999);
Reply to comment: D.S. Wiersma, J. G\'{o}mez
Rivas, P. Bartolini, A. Lagendijk, and R. Righini \textit{Nature}
\textbf{398,} 207 (1999).

\bibitem{gomez99}
J. G\'{o}mez Rivas, R. Sprik, C.M. Soukoulis, K. Busch, and A. Lagendijk,
\textit{Europhys. Lett.} \textbf{48,} 22 (1999);
F.J.P. Schuurmans, M. Megens, D. Vanmaekelbergh, and A.Lagendijk,
\textit{Phys. Rev. Lett.} \textbf{83,} 2183 (1999);
J. G\'{o}mez Rivas, R. Sprik, A. Lagendijk, L.D. Noordam, and C.W. Rella,
\textit{Phys. Rev. E} \textbf{63,} 046613 (2001).

\bibitem{chabanov00}
A.A. Chabanov, M. Stoytchev, A.Z. Genack,
\textit{Nature} \textbf{404,} 850 (2000).

\bibitem{albada85}
M.P. van Albada and A. Lagendijk,
\textit{Phys. Rev. Lett.} \textbf{55,} 2692 (1985);
P.-E. Wolf and G. Maret,
\textit{Phys. Rev. Lett.} \textbf{55,} 2696 (1985);
E. Akkermans, P. E. Wolf, and R. Maynard,
\textit{Phys. Rev. Lett.} \textbf{56,} 1471 (1986).

\bibitem{albada90}
M.P. Van Albada, J.F. de Boer, and A. Lagendijk, \textit{Phys.
Rev. Lett.} \textbf{64,} 2787 (1990)

\bibitem{scheffold97}
F. Scheffold and G. Maret,
\textit{Phys. Rev. Lett.} \textbf{81,} 5800 (1998);
F.Scheffold, W.H\"{a}rtl, G. Maret, and E. Matijevi\'{c},
\textit{Phys. Rev. B} \textbf{56,} 10942 (1997).

\bibitem{labeyrie99}
 G. Labeyrie, F. de Tomasi, J.-C. Bernard, C.A. M\"{u}ller, Ch. Miniatura and R. Kaiser,
\textit{Phys. Rev. Lett.} \textbf{83,} 5266 (1999);
T. Jonckheere, C.A. M\"{u}ller, R. Kaiser, Ch. Miniatura and D. Delande,
\textit{ibid.} \textbf{85,} 4269 (2000).

\bibitem{lodahl05}
P. Lodahl and A. Lagendijk, \textit{Phys. Rev. Lett.} \textbf{94,}
153905 (2005).

\bibitem{patra99}
M. Patra and C. W. J. Beenakker, \textit{Phys. Rev. A}
\textbf{60,} 4059 (1999); \textit{ibid.} \textbf{61,} 063805
(2000); C. W. J. Beenakker, M. Patra, and P. W. Brouwer,
\textit{ibid.} \textbf{61} 051801 (2000); M. Kindermann, Yu. V.
Nazarov, and C.W.J. Beenakker, \textit{Phys. Rev. Lett.}
\textbf{88,} 063601 (2002); P. Lodahl, A.P. Mosk, and A.
Lagendijk, \textit{Phys. Rev. Lett.} \textbf{95,} 173901 (2005).

\bibitem{mandel95}
L. Mandel and E. Wolf,
\textit{Optical Coherence and Quantum Optics}
(Cambridge Univ. Press, 1995).

\bibitem{einstein05}
A. Einstein,
\textit{Ann. Phys.} \textbf{17,} 132 (1905).

\bibitem{cao01} H. Cao, Y. Ling, J.Y Xu, C.Q. Cao, P. Kumar, \textit{Phys. Rev.
Lett.} \textbf{86,} 4524 (2001); M. Patra, \textit{Phys. Rev. A}
\textbf{65,} 043809 (2002); L. Florescu and Sajeev John,
\textit{Phys. Rev. Lett.} \textbf{93,} 013602 (2004).

\bibitem{levitov96}
C.W.J. Beenakker and M. B\"{u}ttiker, \textit{Phys. Rev. B}
\textbf{46,} R1889 (1992); L.S. Levitov, H. Lee, and G.B. Lesovik,
\textit{J. Math. Phys} \textbf{37,} 4845 (1996); D.A. Bagrets,
\textit{Phys. Rev. Lett.} \textbf{93,} 236803 (2004).

\bibitem{johnson66}
F.A. Johnson, R. Jones, T.P. McLean, and E.R. Pike,
\textit{Phys. Rev. Lett.} \textbf{16,} 589 (1966).

\bibitem{schatzel90}K. Sch\"atzel, \textit{Quantum Opt.} \textbf{2,} 287-305
(1990).

\bibitem{deBoer94}J.F. de Boer, M. C. W. van Rossum, M. P. van Albada, Th. M.
Nieuwenhuizen and A. Lagendijk, \textit{Phys. Rev. Lett}
\textbf{73,} 2567 (1994).

\bibitem{rossum99}
M.C.W. van Rossum, J.F. de Boer, and Th.M. Nieuwenhuizen,
\textit{Phys. Rev. E} \textbf{52,} 2053 (1995); M.C.W. van Rossum
and Th.M. Nieuwenhuizen, \textit{Rev. Mod. Phys.} \textbf{71,} 313
(1999).

\bibitem{berk94}
R. Berkovits and S. Feng,
\textit{Phys. Rep.} \textbf{238,} 135 (1994).

\bibitem{sandor06}
B. Sandor, S.E. Skipetrov and F. Scheffold,  in preparation.

\end{thebibliography}
\end{document}